\documentclass{appolb}
\usepackage[T1]{fontenc}
\usepackage[intlimits,sumlimits]{amsmath}
\usepackage{amssymb}
\usepackage[utf8]{inputenc}
\usepackage{hyperref}
\usepackage{xcolor}
\usepackage{graphicx}
\usepackage{listings}
\usepackage{fvextra}

\DeclareMathOperator{\eml}{eml}
\DeclareMathOperator{\edl}{edl}
\DeclareMathOperator{\lde}{lde}
\DeclareMathOperator{\pli}{pli}
\DeclareMathOperator{\plm}{plm}

\begin{document}

\title{Diversity of EML-type operators
}
\author{A. Odrzywołek
\address{Institute of Theoretical Physics, Jagiellonian University,\\
\L{}ojasiewicza 11, 30-348 Kraków, Poland}
}
\maketitle

\begin{abstract}
The discovery of the EML operator, sufficient to evaluate the standard explicit purely transcendental elementary functions, 
has led to considerable interest and discussion across multiple scientific disciplines. However, most authors have focused on the binary EML itself, while numerous similar variants with slightly different properties are now known. This article attempts to close this gap by enumerating and classifying them. We also take this opportunity to clarify common misconceptions related to the EML operator. The principal goal, symbolic regression within an architecture as close as possible to proven neural networks which combine matrix multiplication with a single univariate non-linear activation function, remains beyond reach. Instead, we propose a Möbius layer, with rational functions replacing matrix operations, and showcase the recently discovered activation function eml(x,1/x), which allows exp(x) and ln(x) to be recovered separately, and hence all elementary functions to be evaluated within a rational generalization of the neural network. 
\end{abstract}

  
\section{Introduction and follow-up studies}

The EML operator
\begin{equation}
\label{eml}
\eml(x,y) = \exp{(x)} - \ln{(y)}
\end{equation}
has been proposed in \cite{EML}. It was verified extensively both numerically (including in arbitrary precision) and with a Computer Algebra System (Wolfram Mathematica). The core claim is that using the EML operator, together with input variables and the special distinguished constant 1, one can derive formulas which reconstruct all standard scientific calculator operations. These include the usual arithmetic ($+,-,\times, \div$), $\exp$ and $\ln$, binary exponentiation and logarithm, integers and rationals, constants like $\pi, e$ and $i$ and all trigonometric/hyperbolic functions with their inverses. These formulas work in their respective real domains, except at the endpoints in some cases. For trigonometric functions and $\pi$, the EML operator must use complex intermediate values. This is done entirely within the principal logarithm branch (the fundamental strip of the complex plane). The shortest formulas are obtained using the extended real axis, $\ln{0} = -\infty$ and $e^{-\infty}=0$. Since then, formalized Lean4 proofs \cite{EML-formalization} have appeared supporting the reconstruction, with qualifications discussed in Subsect.~\ref{real_vs_complex}. In what follows we assume the validity of the main result of \cite{EML}. The operator now has its own entry in MathWorld \cite{MathWorldEML}.

The surprisingly simple properties of Eq.~\eqref{eml} sparked fierce internet discussions\footnote{See e.g. the Hacker News thread \url{https://news.ycombinator.com/item?id=47746610} (858 points, 298 comments; accessed August 2026).} about its possible applications and its relation to the logical Sheffer stroke \cite{Sheffer1913} (NAND/NOR binary operators) which enables the computation of all Boolean functions. Indeed, the context-free grammar is  extremely simple 
\begin{equation}
S\!\to\!1\mid x \mid \eml(S,S)
\end{equation}
and is isomorphic to binary trees, one of the best-studied structures in computer science.

One possible application of EML operators and their relatives is symbolic regression (SR). SR attempts to fit data with free-form expressions composed of arbitrary functions and operations. Usually, this requires mixing discrete and continuous approaches \cite{miles_cranmer_2026_21996566}. First, genetic algorithms generate multi-parameter candidate formulas. Then, the free parameters are optimized using e.g. gradient descent. EML, eq.~\eqref{eml}, in principle allows the discrete step to be skipped entirely by using the master formula \cite{EML} encoding a full binary expression tree of depth $d$
\begin{equation}
\label{master}
T_{\eml,d}
\left(
x;
\{a_1,\ldots,a_{k}\},
\{b_1,\ldots,b_{k}\},
\{c_1,\ldots,c_{k}\}
\right),
\; k=2^{d+1}-2,
\end{equation} 
intentionally written in a form resembling the Gauss hypergeometric $_2F_1$ function.
A proof of concept for gradient-based symbolic regression has already been presented in \cite{EML}. However, depth and symbolic recovery rates were very limited. In the follow-up study \cite{2026arXiv260423256G}, three depth-3 architectures were compared on over 12,000 training runs; recovery of the same target ranged from 0\% to 100\% depending on the architecture and training protocol, and balanced (non-chain) trees were never recovered. In these experiments, the optimization landscape, not expressivity, is therefore the limiting factor. This is precisely where we hope EML variants (Sect.~\ref{variants}, Table~\ref{StachowiakForm}) might help.

EML trees were used as building blocks for simultaneous function and antiderivative discovery \cite{belaiche2026additiveatomicforestssymbolic}, as response modules in reduced models of biological dynamics \cite{2026arXiv260502972E}, and as interpretable edge mechanisms in causal structure learning \cite{asanuma2026emlcdcausalmechanismrecovery}. A recent theorem also establishes universal approximation by EML trees in $W^{k,\infty}$, together with a practical fitting procedure \cite{Germany2026Universal}. Independent implementations now exist in Rust, Python, and JavaScript
\cite{EML-implementations}.

The rest of the article is organized as follows. In Sect.~\ref{polemika} we address some questions which arose after the EML discovery was made public. Sect.~\ref{variants} showcases the known variants of the EML operator. In Sect.~\ref{mobius} we propose a new architecture upgrading neural networks, and the article is concluded with a short summary.

\section{Relevant mathematical background and common misconceptions \label{polemika} }

Months of intense discussion of the EML operator have resulted in a much deeper understanding of its relation to prior mathematical knowledge. While EML and some of its variants, were discovered by an exhaustive brute-force numerical sieve, more explainable approaches now exist \cite{Stachowiak}. Some new formulas are far more complex than a direct search using limited compute could reach.

In the following subsections we take the opportunity to explain common misconceptions and \emph{nitty-gritty details}, which resurfaced during discussions of the EML operator. 

\subsection{All elementary functions?}

Some confusion comes from the not universally accepted notion of \textit{elementary functions}. The meaning of this phrase is different not only between scientific disciplines, but also between languages and countries. For example, in Poland, math teaching clearly distinguishes between \emph{explicit} and \emph{implicit} ones (pl. funkcje jawne/uwikłane). While in \cite{EML} unambiguous enumerative definition of what is meant by \textit{elementary functions}, i.e. explicit finite expressions composed using standard scientific calculator buttons, was provided, some confusion among unprepared readers arises. Our definition is the same as in complex analysis textbooks, see e.g. \cite[Chapter 5, Elementary Functions]{complexanalysisorg}; see also \cite[p.~746]{MityushevRogosin2020}. However, some readers of \cite{EML} didn't follow further than reading the preprint title, not to mention its abstract or full text, and just a few reached the Supplementary Information.  The purely mathematical definition is used mainly in differential algebra, and arose in the context of integration in finite terms. This is a historically famous problem, second only to the solution of cubics/quartics/quintics in importance for shaping the modern notion of what we mean by elementary functions. The well-known example providing a rationale for the mathematical definition of \emph{elementary function} is the symbolic integration of rational functions such as $\int 1/(x^4+1) dx$. The classic algorithm  for that is at the core of the STEM teaching curriculum, and uses decomposition of the rational integrand into the partial fractions. To achieve this, the roots of the denominator are required to be known. If we insisted that they must be found in the form of radicals,  then a generic rational function with fifth or higher degree in the denominator would be not be symbolically integrable in finite terms. Therefore, solutions to all polynomial equations are \textbf{adjoined} (using differential algebra terminology). To the great surprise of many STEM practitioners outside pure math, solutions to e.g. $x^5 - x + z =0$ with unknown $x$ and parameter $z$ are therefore (e.g. the 2nd real root) \textit{elementary functions} $f(z)$ of the variable $z$ in the above sense, despite being explicit only after using a hypergeometric function which does not reduce to exp-log expressions \cite{Beukers2014Hypergeometric}
$$
f(z) = z \; {_4 F _3} 
\left( 
\{ \tfrac{1}{5},\tfrac{2}{5},\tfrac{3}{5},\tfrac{4}{5} \}; \{ \tfrac{1}{2},\tfrac{3}{4},\tfrac{5}{4} \};
3125 z^4/256 
\right).
$$
 So if one interpreted title of \cite{EML} in a narrow differential-algebraic strictly mathematical sense, then EML expression would provide e.g. constructive exp-log solution $f(a,b,c)$ to Hilbert's 13th problem, which involves roots of the 7-th order polynomial $  x^7 + a\, x^3 + b\, x^2 + c \, x +1 =0$ with 3 parameters $a,b,c$. This is, of course, not what \cite{EML} claims, and the algebraic-function version (cf. Subsect.~\ref{prior}) of the Hilbert's 13th remains open.

The above didn't stop some authors from proving that EML is not able to express ''ALL'' \emph{elementary functions}, despite the  difference in definitions of this phrase. An example of a constant which cannot be expressed in terms of the EML operator and 1 is Chaitin's constant \cite{Carney2026}. However, it is also not computable. Another blog post discusses the roots of the generic quintic \cite{notall}, which again is not an EML-class expression. What is expressible by EML are actually Chow's \cite{Chow1999} elementary numbers. Carney presents the equality of these sets as Proposition~1 in \cite{Carney2026}.

As a bottom line I would like to stress that choosing the title for the EML paper \cite{EML} was not an easy choice. To include all relevant information  in the title, it would have to be ridiculously long and complicated, e.g. 
\begin{center}
\emph{A single binary operator for computing exp-log functions on the extended real line, almost everywhere, under IEEE 754 semantics, with algebraic adjunctions replaced by Chow's elementary numbers, via complex-valued intermediate values, from the constant 1?}
\end{center}
or similar. This would be consistent with the modern struggle to attract the attention of potential readers, who only skim titles, rarely read the abstract, and do not even attempt to read the full text, not to mention the \textit{Supplementary Information}. However,  information about the existence of the EML operator might be important to the large community of machine learning and electronic engineers, bioinformaticians, theoretical computer science experts and educators. A title narrowly aimed at mathematicians would mislead the core audience. On the other hand, a casual title, e.g., \emph{Two-button calculator} would suggest recreational math, without any applications in sight. For most STEM educated people around the world, elementary functions = scientific calculator operations. Period.

\subsection{\label{real_vs_complex} Complex domain and branches}

Another common misconception related to the EML operator is the use of complex intermediate values to generate real functions. While this sounds natural for theoretical physicists, e.g. quantum mechanics operates exactly that way (complex wave function, real measurements), more mathematically aligned people were confused: what is the domain of the \eqref{eml}? If one presumes it is the natural real domain $\mathbb{R}\times\mathbb{R}^+ \to \mathbb{R}$, the entire reconstruction becomes impossible. As proved by Hardy in \cite{Hardy1912} one cannot obtain $\sin{x}$ from $\exp, \ln$ and arithmetic within the real domain, because it oscillates at infinity \cite[Chap.~III, Sect.~2, Theorem on p.~18]{Hardy1910Orders}.
If not real, then maybe a complex mapping $\mathbb{C}^2 \to \mathbb{C}$ of EML arguments is enough for the full reconstruction of the elementary function chain? However, while $\exp(x)$ is well-defined on the entire complex plane, $\ln(y)$ is infinitely-valued. It is a true function only when defined on its Riemann surface, except at (complex) zero. In practice, $\eml(x,y)$ is defined on $\mathbb{C} \times (\mathbb{C}\setminus\{0\})$ using the principal branch of the complex Log. 
Moreover, all functions to be reconstructed by EML are \emph{real}, $\mathbb{R}^n \to \mathbb{R}$. Therefore, complex inputs to $\eml(x,y)$ appear only at internal expression nodes. Both inputs and outputs of EML-expressions are real, while inputs are additionally restricted to be within the real domain of the reconstructed expression, with the possible exception of domain endpoints \cite{EML-formalization}.

The main surprise of \cite{EML} preprint result is that the above \emph{do work}. In principle, one would expect that for some big EML expression tree internal values escape fundamental strip, and entire reconstruction breaks. Initially, a lot of attempts were made to find such a counterexample. All of them appeared to be bugs in EML compiler, or even Wolfram Mathematica \pmb{FullSimplify} procedure failures claiming false formulas to be true. Formal Lean4 results are available in \cite{EML-formalization}, although some trigonometric statements use input-dependent expressions or real/imaginary-part projections.

Note that question whether similar reconstruction for $\mathbb{C} \to \mathbb{C}$ elementary functions, even reduced to principal branch, is possible, is separate and unanswered question. If it is possible, it likely would require operator much different from \eqref{eml}.
Reconstruction of complex-valued multivalued elementary functions in similar manner looks impossible task, as they are defined on distinct Riemann surfaces.

\subsection{Use of extended real axis and logarithm at zero}

The original construction used in \cite{EML} uses a Bourbaki-style extended domain, including infinite points. In particular, the reconstruction relies on the formulas
$$
e^{-\infty} = 0, \qquad \ln{0} = - \infty. 
$$
This is confusing for many STEM practitioners, for whom the logarithm at zero is undefined, and $\pm \infty$ are not valid arguments or values. Moreover, since EML operates in the complex domain, one must be careful whether we are using complex infinity, directed infinity, or real infinities. In practice, the above poses no problem, as we use EML expression within Computer Algebra Systems like Mathematica, which handle the above natively. The same applies to IEEE754 numerical floating-point computations, which define and handle infinities automatically. Therefore, depending on the level of error handling, EML expressions are easily evaluated in many programming languages. But not in all of them. For example, Python's and Maple's standard logarithm routine raises an error at zero. This motivated the creation of \emph{clean} EML compiler, which never touches the troublesome $\ln(0)$, at the expense of much longer expressions. So far this looks doable. In particular, Lean4 proofs must avoid the complex Log(0), because, due to the
requirement for Log to be a total function (full $\mathbb{C}$ domain, including zero) ''junk'' value Log(0)=0 is imposed. This breaks reconstructions that use the extended value at zero, and \cite{EML-formalization} must avoid it.

\subsection{EML as NAND/NOR equivalent for continuous math}

As A.~Rieu pointed out \cite{Rieu2026NORNAND}, the history is somewhat more intricate than the modern identification
of the Sheffer stroke with NAND suggests. Peirce discovered the functional completeness of both NAND and NOR around 1880, although
his result remained unpublished, and Stamm published both forms in 1911. Sheffer's 1913 paper selected the NOR interpretation
(``neither-nor''), whereas Nicod's 1917 treatment adopted the NAND interpretation now conventionally associated with the stroke
\cite{Stamm1911,Sheffer1913,Nicod1917}. Their later significance for digital computation rests especially on Shannon's identification of
Boolean algebra with relay and switching-circuit synthesis \cite{Shannon1938Switching}.

In some sense the situation is similar to EML. While \eqref{eml} was discovered first, at least five equivalents are now known, cf. Table~\ref{StachowiakForm}. The notion of ''sole sufficient operator'' in both cases (NAND, EML) in fact refers to multiple cases: NAND/NOR \textit{vs} EML/EDL/LDE/PLI/PLM. The analogy between digital NAND and continuous EML is quite close. One difference is the requirement to use the constant 1 for EML to work, while NAND does not require this. But in practice, all circuits use external 0/1 inputs anyway. NAND can compute any Boolean function and approximate any other function. The same is true for EML. It can compute any elementary function, which can in turn approximate any functions, using polynomials (series, Chebyshev) rational functions (Pade), Fourier series, integration (Gaussian, double-exponential), ODEs (Runge-Kutta methods) and last but not least neural networks. All aforementioned methods have proved useful in science.

The problem of generating EML values (output) is technical, not fundamental. In the late XIX and early XX century no one had any idea how to implement NAND/NOR in a massive, efficient way. Babbage's Analytical Engine \cite{Bromley1982AnalyticalEngine} was purely mechanical, Shannon \cite{Shannon1938Switching} discussed electromechanical switches. Only in the 60s, first in the Apollo Guidance Computer, did we start using integrated circuit NOR and later NAND TTL gates. We hope the engineering and adoption of EML-type operators will be much faster, in either digital or analog form.

\subsection{Use of the brute-force search}

Exhaustive computational searches have a long and successful history in
the discovery of mathematical counterexamples. For example, Lander and
Parkin used a direct search to disprove Euler's sum-of-powers
conjecture by finding
$
  27^5+84^5+110^5+133^5=144^5
$
\cite{LanderParkin1966}. Another example is the exhaustive enumeration by
Dokovic showing that Williamson matrices of order 35 do not exist,
thereby disproving the Williamson conjecture
\cite{Djokovic1993}. Computational searches have also been used to formulate
new conjectures \cite{BorweinBailey2008}. Here we used them to select arithmetic operators with given properties. Therefore, enumeration was used recursively. First, we enumerated candidate operators, then all formulas composed of them. A common misconception related to \cite{EML} is that this was all that was done. In fact, the exhaustive search was meant to provide \emph{candidate} operators. EML itself, once marked as first possible successful candidate, was subject to very extensive numerical and symbolic hand-crafted verification.

By design, the fast numerical sieve, based on numerical constant recognition, might return false-positives. This can happen due to numerical round-off errors or floating-point tautologies. We recall that floating-point numbers, while intended to approximate reals, in fact are dyadic rationals. Their set is huge, but finite, and calculations are discrete, not continuous. Our procedure works as follows. First, we generate some elementary formula which is representing two-input operator, similar to \eqref{eml}. Then we want to know, if some combination of this operators is equivalent to standard mathematical operation, like square root or multiplication. Instead of slow and error prone symbolic simplification, we attempt to establish floating-point identity at some point. This cannot be value like, say $x=-2/3, y=
\pi$, because both of these numbers are already elementary. We need some truly transcendental constant. The best situation would be if we know that it is at least irrational, but for most convenient cases, like Khinchin $K\simeq2.68545$, Glaisher $A\simeq1.28243$, Euler gamma $\gamma \simeq 0.577216$ or Catalan $C=0.915966$ constant there are no relevant proofs\footnote{See, however, recent preprint https://arxiv.org/abs/2609.04176 on Catalan.}. Later we proceed as if these constants were truly outside the exp-log class. Once we compute numerically a high-precision value, e.g. $\sqrt{A} \simeq 1.1324429915455447052953341751903$ or $K \times C \simeq 2.4597816377901852095513549038665$, we can run constant recognition software 
\cite{SrokaCR} on the result. If the fast numerical search provide an equivalent candidate formula using e.g. $\eml(x,y)$ and 1 only, which agree to some assumed error of a few ULPs with the expression sought, then additional checks (arbitrary precision, symbolic, multi-point) are executed.
If they pass, the formula is marked a valid candidate, and we proceed to the next case \cite{EML,EML-formalization}.

If the entire set of operations which define our ''elementary functions'' is reconstructed in the above way, the result is subject to detailed, hand-crafted symbolic and numeric verification. For details, see \cite{EML}.

What might be surprising is that the above procedure was able to discover anything. One might call it luck, however it was the culmination of very long research on brute-force methods in symbolic regression, started as early as 2009\footnote{Cf. e.g. \url{https://th.if.uj.edu.pl/\~odrzywolek/homepage/presentations/PL/Approx.pdf} (in Polish) where the KAN idea \cite{KAN} was presented as well.}. The idea was to use a mix of random/genetic and exhaustive search to sweep unsolved problems at scale. In practice it was limited to occasional solving algebraic equations, evaluating definite integrals, solving ODEs or identifying numerical constants from various sources of experimental mathematics or physics. To our unpleasant surprise, for a decade this programme led only to the re-discovery of already known results in different forms. See for example \cite{1625336}.
This \emph{unreasonable effectiveness of human science} in solving problems was a surprise. It looked like everything that could be found, was already found. No gaps in knowledge, no missing solutions, no forgotten old problems. In fact, the EML discovery was the first major success of the above philosophy. In some sense, finally it was demonstrated to be right. It is conceivable that throwing substantial computational power and programming effort at some unsolved scientific  issues could work in the above way.


Now, in 2026, deep learning neural networks in the form of LLM chatbots (ChatGPT, Grok, Gemini) and programming agents (Codex, Claude Code) are overshadowing any other attempts in machine learning. In the last few weeks, torrents of counterexamples\footnote{May 20, 2026: internal OpenAI model, Erdős unit-distance conjecture (1946) \cite{OpenAIUnitDistance2026}; July 19: L.~Alpöge, Jacobian conjecture (Keller, 1939), false for $n \geq 3$ \cite{Gao2026Jacobian}; July 22: D.~Rybin with GPT-5.6 Pro, Dinitz–Garg–Goemans conjecture (1999); July 29: Arathoon, Ball, Kvalheim with GPT-5.6 Sol, Maxwell conjecture (1873) on equilibria of point charges \cite{Arathoon2026Maxwell}; Aug 1: OpenAI ``Ten advances'', incl. first non-sofic group and disproof of Connes's rigidity conjecture \cite{OpenAITenAdvances2026}.} to decades-old old conjectures (\emph{mensis mirabilis}) generated by those AI systems have left the world's top-class mathematicians, including Fields medalists, in shock. The broader debate is reflected in the earlier Leiden Declaration \cite{LeidenDeclaration2026}. Some of these conjectures are iconic in modern math, and many geniuses attempted solving them. In retrospective, some of these conjectures (e.g. the Jacobian conjecture, DGG, Maxwell) could have been disproved many years ago by brute force search, but no one tried hard enough. Things are happening too fast now to even think about possible consequences. But it looks like \emph{bitter lesson} \cite{Sutton2019} i.e. throwing an insane amount of computing power and data at any optimization problem, is working nearly as good as the most sophisticated dedicated ''intelligent'' solutions, including human science. This in turn raises the major question of the early XXI century: \emph{what is intelligence?}

Brute-force exhaustive enumerative search seems at first glance to be the opposite of intelligence. The iconic use case of a student who patiently substitute subsequent integers\footnote{For years, we benchmarked the ability of hardware to substitute ALL $2^{32} \simeq 4.3 \times 10^9$ single-precision floats into transcendental equation $e^x=x^2$ in an attempt to solve it. Initially, around 2004, a single-core Athlon XP required 15 minutes to complete this task. Modern multi-core CPUs are at the seconds level. Now, an RTX6000 PRO using CUDA completes the search in \ldots 10 ms!} into a homework equation in an attempt to solve it is an example of nonsense work. We expect them to use some algorithm instead. But where did the algorithm come from? We can enumerate candidate algorithms as well. Once found, it can be used forever, and possible enhancing future brute-force search for other objects. This looks more like developmental biology than scientific progress. The major question of our times is therefore not what intelligence is, but where genuinely new knowledge comes from. In our opinion, it is always brute-force search in disguise. Justification and explanation for discovery is usually provided \emph{post-factum}. In this light, there is no surprise that modern AI systems are said to be unable to create new ideas beyond mixing and application of what was in the training data. The fact that ''trivial'' counterexamples to famous conjectures are now found by AI is then evidence of the negligence of the scientific community in the valuation of brute-force methods rather than a symptoms of emerging AGI (Artificial General Intelligence) or ASI (Artificial Super Intelligence).

This leaves the intriguing possibility, that redirecting the world's compute into exhaustive search instead of AI training might be the most efficient use of it. Notable examples of such searches of the ''mathematical universe'' \cite{TegmarkMathematicalUniverse} include Wolfram's Physics Project \cite{Wolfram2020} and Taelin's discrete program search on the HVM \cite{HVM}.

\subsection{Relation to Turing machine and universal computation}

A frequent question is how EML relates to established computational models like the Turing machine. EML real-function formalism is closer in spirit to analog computations \cite[Chap.~12, Sec.~4, note (c), ``Continuous computation,''
p.~1128]{NKS}, in which the exponential function is a primitive object. Even integers must be constructed from it. A list of the first 100 integers expressed in pure EML form is given in \cite{EMLCodeGolf}. A more striking example is the representation of $\pi$. Its decimal expansion $\pi=3.14159\ldots$ has been computed to 314 trillion digits, turning it into a big data object \cite{PiRossi}. In contrast, EML gives the following  \emph{exact finite expression} \cite{MaplePrimes}:
\begin{Verbatim}[
  breaklines=true,
  breakanywhere=false,
  breakafter={,},
  breaksymbolleft={},
  breaksymbolright={},
  breaksymbolindentleftnchars=0,
  breakaftersymbolpre={},
  breakaftersymbolpost={},
  fontsize=\footnotesize
]
eml(eml(eml(1,eml(eml(1,eml(1,eml(eml(1,eml(eml(1,eml(eml(1,eml(1,eml(eml(1,1),1))),1)),eml(eml(eml(eml(eml(1,eml(eml(1,eml(1,eml(eml(1,eml(1,eml(eml(1,eml(eml(1,eml(eml(1,eml(1,eml(eml(1,1),1))),1)),eml(1,1))),1))),1))),1)),eml(eml(eml(1,eml(eml(1,eml(1,eml(eml(1,1),1))),1)),eml(eml(1,eml(eml(1,eml(eml(eml(1,eml(eml(1,eml(1,eml(eml(1,1),1))),1)),eml(eml(1,eml(eml(1,eml(eml(1,eml(eml(1,1),1)),eml(eml(eml(1,eml(eml(1,eml(1,eml(eml(1,1),1))),1)),eml(1,1)),1))),1)),1)),1)),1)),1)),1)),1),1),1))),1))),1)),eml(eml(eml(1,eml(eml(1,eml(1,eml(eml(1,1),1))),1)),eml(eml(1,eml(eml(1,eml(1,eml(eml(1,eml(eml(1,eml(eml(1,eml(1,eml(eml(1,1),1))),1)),eml(1,1))),1))),1)),1)),1)),1).

\end{Verbatim}

This is confusing for many computer scientists, who reason in the following way. First, we need a computationally intensive binary algorithm for the evaluation of $e^x$, then we define \eqref{eml} to compute \ldots $e^x$ again. This looks circular. It makes no sense until we have a method to evaluate $e^x$ directly. Physics, however, provides exponentials directly in numerous processes: light attenuation, exponential growth, RC circuit, radioactivity $T_{1/2}\!=\!\tau \ln{2}$, statistic $e^{-\frac{E}{k_B T}}$, entropy $S\! =\! k_B \log{W}$, cosmological inflation $e^{H_\infty t}$, the Gaussian curve $e^{-x^2}$, quantum phase $e^{i \varphi} |\psi \rangle$ etc. Evidently, the existing digital architecture is unsuitable for EML computations. An FPGA implementation may be useful in specific conditions \cite{Taylor2026EMLFPGA}, namely limited resources, e.g. on micro satellites, but still must use a standard discrete algorithm to compute $e^x$. For efficiency, we need some unconventional, non-von Neumann architecture. A step in the right direction is provided e.g. by \cite{Extropic}.

\subsection{Prior knowledge \label{prior}}

While the discovery of a simple analytical single binary operator for the computations of elementary function was a true surprise, and the majority\footnote{This also included all top AI LLM models trained with a cutoff before April 2026, as such knowledge was not in their training data, and was not an obvious consequence of it.} before April~2026 would convincingly state that this was impossible, some hints of its existence were hidden in plain sight long before that.

Mathematicians attempted to answer related questions. In rational function theory it was probably known \cite{Hamkins2011UnifyOperations} that rational functions can be generated over the field of rational numbers $\mathbb{Q}$ using a single operator
\begin{equation}
\label{hash}
x \# y  \equiv \frac{1}{x-y}
\end{equation}
plus an unclear number of terminal constants. We found that this is indeed possible using both 0 and 1 as distinguished constants\footnote{In fact, the constant 1 can be replaced by any non-zero rational.}. In this sense, for rational functions with integer coefficients, e.g., $11(x^2+x+3)/(x^4-4)$, the ''hash operator'' \eqref{hash} is indeed an analogue of the EML operator. However, it is not clear how to extend \eqref{hash} to exp-log functions in any other way than by extending the hash operator with $\exp(x)$ and $\ln(x)$ themselves. But if we allow for this, then plain subtraction is already good enough, cf. Table~\ref{StachowiakForm} or Calc~2 from Table~2 in \cite{EML}. Moreover, no distinguished terminal constant is required. The statement that the ''hash'' operator is equivalent to EML is simply false.

Another piece of knowledge missed by \cite{EML} is Hua's identity \cite{Hua1949}, valid for any \emph{division ring}, hence also for real/complex numbers. For real/complex numbers, Hua's identity allows one to express multiplication using addition/subtraction and inverse
\begin{equation}
\label{hua}
a b a  = a - \left[ a^{-1} + \left(b^{-1} - a\right)^{-1} \right]^{-1}.
\end{equation}
The Kolmogorov-style RPN complexity of the above formula is too large for exhaustive search, unless we allow for the use of the reduced mass 
$
u || v = (u^{-1} + v^{-1})^{-1} 
$
(parallel sum) as a primitive arithmetic operation. Then the reciprocal can be computed using $a^{-1} = 1 - [1 || (a-1)]$, identity \eqref{hua} gives multiplication, and the EML reduction of the operator count using the constant 1 sounds less mysterious.

In the abstract theory of binary operators on arbitrary sets, a construction reducing any number of binary operators to a single one is known \cite{Goldstern}. Here we present an example construction. The reader is encouraged to compare it to the elegance of the EML operator. In the following, we reduce the four basic operations ($+, -, \times, /$) to a single ''star'' operator using four distinguished integer constants: $-1,-2,-3,-4$. First, we create four ''compactified'' copies of the real line
$$
q_k(x) = 10 + k + \frac{1}{2} + \frac{\arctan{x}}{\pi}.
$$
The inverse operation is, of course
$$
Q_k(z) = \tan{\left[ \pi \left(z - 10 - k -1/2 \right) \right]}.
$$

Now we define a single operator as
\begin{equation}
\label{4const}
x \ast y = 
\begin{cases}
q_0(y) & \text{for} \; x=-1,\\
q_1(y) & \text{for} \; x=-2,\\
q_2(y) & \text{for} \; x=-3,\\
q_3(y) & \text{for} \; x=-4,\\
Q_0(x) + y      & \text{for} \; 10 < x < 11,\\
Q_1(x) - y      & \text{for} \; 11 < x < 12,\\
Q_2(x) \times y & \text{for} \; 12 < x < 13,\\
Q_3(x) / y      & \text{for} \; 13 < x < 14.\\
\end{cases}
\end{equation}
It is a simple exercise to verify that we have
\begin{eqnarray*}
x + y       = [(-1) \ast x] \ast y, \\
x - y       = [(-2) \ast x] \ast y, \\
x \times y  = [(-3) \ast x] \ast y, \\
\frac{x}{y} = [(-4) \ast x] \ast y. 
\end{eqnarray*}

Because $[(-1)\ast(-1)]\ast (-1) = -2$ etc., one can reduce the number of terminal constants to one. Moreover, we have verified in Mathematica that another implementation of the \cite{Goldstern} idea leads to a single ''diamond'' operator working without any constants, see Appendix~A.

While the procedure \eqref{4const} works for binary operators only, since arbitrary arity reduces to binary by a classical theorem of Sierpiński \cite{Sierpinski1945}, one can extend it to unary exp-log using dummy operators, e.g., $\mathrm{pwr}(x,y)=\exp(x)$ and $\mathrm{lg}(x,y)=\ln{x}$. This affirmatively answers the question raised by \cite{EML} in the set-theoretic reading. But not for analytic operators \cite{Stachowiak}.  

In this sense, \cite{Goldstern} anticipated the existence of a single operator. But a Goldstern-type operator is essentially a set-theoretic multiplexer using a lookup table (see Appendix~A). It is more similar to the Kolmogorov-Arnold bypass of Hilbert's 13th problem, allowing for the reduction of any continuous function of $n$ variables to $2n+1$ univariate ones using an $f$-dependent ''lookup function'' and addition, in its modern form \cite{SprecherTransAMS1965}
$$
f(x_1, \ldots, x_n) = \sum_{q=1}^{2 n+1} g \Biggl( \sum_{p=1}^n \lambda_p \phi_{q}(x_p) \Biggr).
$$
Neither is useful in practice due to pathological mathematical properties. Function $g(x)$ is $f$-dependent, 
only continuous, and highly irregular even for analytic $f$, and the inner functions cannot be chosen smooth. This is in contrast to EML-type operators, which are simple analytical formulas.

\section{Diversity of operators \label{variants}}

\subsection{EML variants}

Since the original discovery of \eqref{eml}, it was quickly realized that at least two similar close cousins exist. The first is EDL (Exp-Divide-Log)
\begin{equation}
\edl(x,y) = \frac{e^x}{\ln{y}},
\end{equation}
paired with the Euler number $e$. Another one is -EML
\begin{equation}
-\eml(y,x) = \ln{(x)} - \exp(y),
\end{equation}
paired with $-\infty$. Once a faster sieve implementation became available, it become clear that the above three are not, like initially believed, related by some Möbius transform. Two new variants of the EDL, LDE (Log-Divide-Exp) were found
\begin{equation}
\lde(x,y) = \frac{\ln{x}}{e^y},
\end{equation}
one working with 0, other with 1.

Other variants are, PLI (Power-Log-Inverse)
\begin{equation}
\pli(x,y) = (\ln{x})^{1/y},
\end{equation}
and PLM (Power-Log-Minus)
\begin{equation}
\plm(x,y) = (\ln{x})^{-y},
\end{equation}
both with the constant 1.

Related constructions of arithmetic operations through bijections occur in non-Newtonian calculus \cite{Czachor2019FractalArithmetic}.
In \cite{Stachowiak}, Stachowiak proposed a general scheme for generating EML-type operators. Here we repeat his definitions 

\begin{equation}
 \label{gen}    
  S(x,y) = M\left( f(x), f^{-1}(y)\right),
\end{equation}

\begin{subequations}
\label{axioms}
\begin{equation}
    \exists_{\hat{e}}  M(x,\hat{e}) = x,
\end{equation}
\begin{equation}
    M(x,x) = \hat{e},
\end{equation}
\begin{equation}
    M(x,M(y,z)) = M(z,M(y,x)).
\end{equation}
\end{subequations}
where $S(x,y)$ is an EML-type binary operator, $f(x)$ -- generating function, $M(x,y)$ -- non-commutative operator,  $\hat{e}$ -- neutral  element, $c=f(\hat{e})$ -- distinguished constant. Noteworthy, all known operators can be cast in Stachowiak's form, cf. Table~\ref{StachowiakForm}. It is however an open problem what kind of pair $M, f$, which defines an EML-type operator, is allowed. Stachowiak discussed the example of $f(x) = \cos{x}$, but was unable to complete the entire elementary function chain. Our brute-force search depth was  too shallow to improve significantly on that. However, using \textit{prosthaphaeresis}, one can compute multiplication using \emph{two} constants, $c=1$ and $a=\arccos(1/4)$ with the same $S=\cos{x} - \arccos{y}$. Reciprocal/division is still unreachable, though.

\begin{table}
\caption{\label{StachowiakForm} Known EML-type operators in Stachowiak's form.}
\begin{tabular}{l||ccccc}
Name & $M(x,y)$        & $\hat{e}$ & $c=f(\hat{e})$ & f                 & $f^{-1}$ \\
\hline
EML  & $x-y$           &  $0$      & 1              & $\exp{x}$         & $\ln{x}$ \\
-EML & $x-y$           &  $0$      & $-\infty$      & $\ln{x}$          & $\exp{x}$ \\
EDL  & $\frac{x}{y}$   &  $1$      & $e$            & $\exp{x}$         & $\ln{x}$ \\
LDE  & $\frac{x}{y}$   &  $1$      & $0$            & $\ln{x}$          & $\exp{x}$ \\
PLI  & $x^{(1/\ln{y})}$&  $e$      & $1$            & $\ln{x}$          & $\exp{x}$ \\
PLM  & $x^{(1/\ln{y})}$&  $e$      & $1$            & $\frac{1}{\ln{x}}$& $e^{1/x}$ \\
     &                 &           &                &                   &            \\
\end{tabular}
\end{table}

The operator $M(x,y)$ is in all cases from Table~\ref{StachowiakForm} the inverse of a Bennett's symmetric (commutative) hyperoperation \cite{Bennett1915} 
$$
F_n(a,b) = \exp^n ( \ln^n a + \ln^n b)
$$
of increasing order:
\begin{enumerate}
\item[$n=0$] addition/subtraction  $F_0(x,y) = x+y \to x-y$ (1st order hyperoperation, i.e. repeated zeration/successor)
\item[$n=1$] multiplication/division $F_1(x,y) = x \times y \to \frac{x}{y}$ (2nd order hyperoperation, i.e. repeated addition)
\item[$n=2$] symmetrized power $F_2(x,y) = x^{\ln{y}} \to x^{1/\ln{y}}$ (commutative analog of the 3rd order hyperoperation, i.e. repeated multiplication)
\end{enumerate}

We must conclude that while the framework proposed by \cite{Stachowiak} gives some order and insight into the structure of EML-type operators, we still lack a general understanding, and exhaustive or fine-tuned search is still the only viable method to find them.

\subsection{Ternary variants}

The requirement for a distinguished constant at the inputs of EML and its variants is troublesome for the implementation of the master formula \eqref{master} and its use for machine learning or symbolic regression. However, the use of this form of the ''switch'' seems unavoidable 
\cite{Hamkins2011UnifyOperations,Goldstern} (see also Subsect.~\ref{prior}). A workaround is to use a ternary operator instead. So far, two \cite{EML} were found:
\begin{subequations}
\label{ternary_T}
\begin{equation}
T_1(x,y,z) = \frac{e^x}{e^y} \times \frac{\ln{x}}{\ln{z}},
\end{equation} 
\begin{equation}
T_2(x,y,z) = \frac{e^x}{e^y} \times \frac{\ln{z}}{\ln{x}}.
\end{equation} 
\end{subequations}

Noteworthy, $T_i(x,x,x)=1$, and no distinguished constant is required to generate all expressions. Unlike the interpolating ternary of \cite{Stachowiak}, \eqref{ternary_T} involve no case distinction.

\section{Möbius layer neural networks \label{mobius}}

All classic neural networks use a single univariate real activation function. The essential question is whether we can find a similar function, which not only approximates data, but is able to generate elementary expressions in exact form as well. Such a construction could connect trainable networks with symbolic regression. Therefore, the answer to the above question is very important for progress in machine learning.

The EML operator \eqref{eml} might be viewed as some variant of 2-input complex-valued activation function, in analogy to sigmoid or ReLU in other variants of computational networks. For EML, the network graph is a parameterized binary tree. As demonstrated in \cite{EML} training such a tree is difficult, and so far has been demonstrated in a limited capacity. Its unique property, however, is the ability to express all elementary functions, which have proved useful in STEM and science over the last centuries. On the opposite side we have machine learning, in the form of deep neural networks. While networks with sigmoid or ReLU activations cannot express functions like $\sin{x}$ exactly in the full real domain, they have proven (both in theory and practice) the ability to approximate any of them. But the decisive property favoring the latter in applications is a working optimization procedure (e.g. Adam, stochastic gradient method) which can be extended to industrial-scale networks with trillions of parameters as of 2026, with no ''wall'' in sight preventing further increase.

Therefore the natural follow-up question, urgent after the discovery of the EML, is whether some of its variants will exhibit the scaling properties of modern deep neural networks, keeping ''symbolic'' capabilities of the EML-trees. Due to the aforementioned theorem by Hardy (see Subsect.~\ref{real_vs_complex}), a neural network with a real exp-log activation is unable to compute e.g. $\sin{x}$. What is the minimal modification then, which would enable exact elementary functions in deep learning, keeping a convenient optimization landscape?  Below we provide some hints as to which direction the search should proceed.

\begin{figure}
\centering
\includegraphics[width=\textwidth]{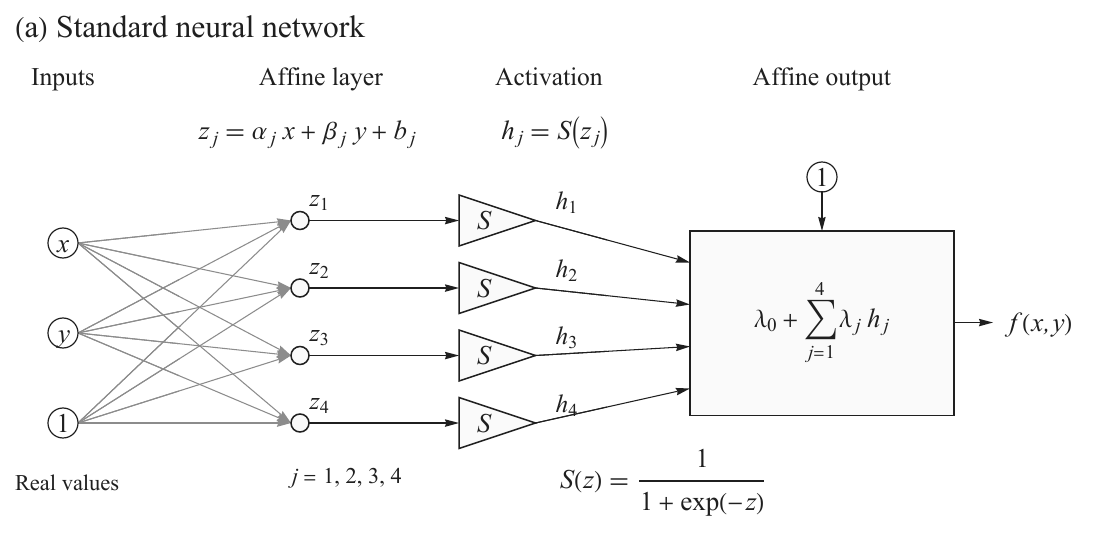}\par\medskip
\includegraphics[width=\textwidth]{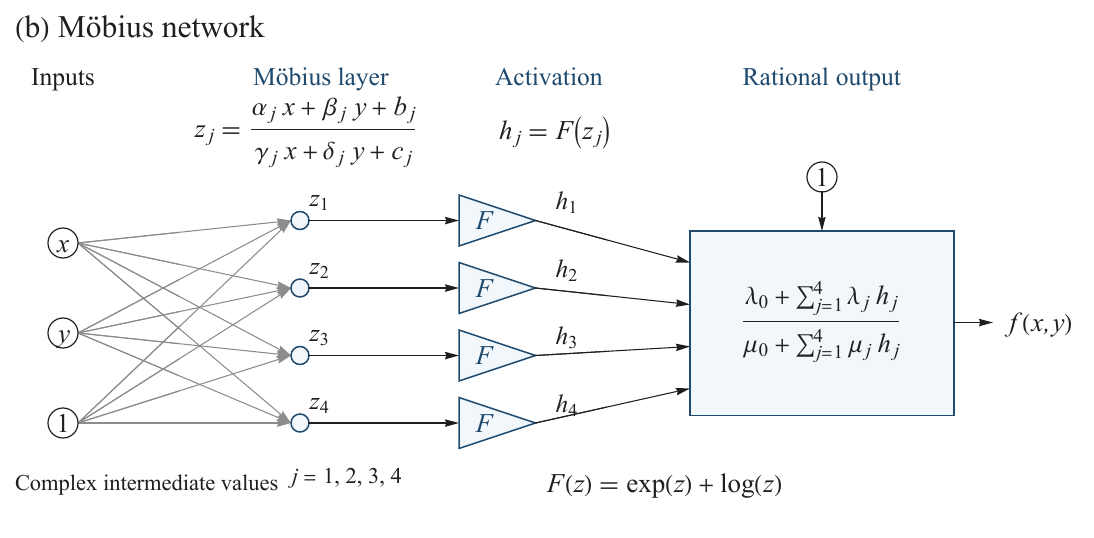}
\caption{\label{arch} Comparison of the standard neural architecture (top) and the proposed generalized ''Möbius'' network (bottom). Both include constant inputs supplying the biases. In the proposed network, the affine input and output maps become ratios of affine forms, and $S$ is replaced by $F$. Each panel shows one layer with four activation units.} 
\end{figure}

A typical neural network rewritten in EML-style language is a ''calculator'' with basic arithmetic operators ($+,-,\times$) (''matrix multiplication'') plus a single univariate non-linear activation function, e.g. the logistic sigmoid
\begin{equation}
\label{sigmoid}
S(x) = \frac{1}{1+e^{-x}},
\end{equation}
known in physics as the Fermi-Dirac distribution. Noteworthy, the set of allowed operations \emph{does not} include division. We now attempt to modify the standard neural network, shown in Fig.~\ref{arch}, top panel, to enable exact elementary functions.

To achieve this goal, we employ three modifications: (i) replace the activation function, (ii) use complex numbers internally, (iii) add division to the allowed arithmetic operations. The new (complex valued) activation function is
\begin{equation}
\label{holly_molly}
F(z) = e^z + \ln{z} \equiv \eml\left( z,z^{-1} \right).
\end{equation}

The crucial identity
\begin{equation}
\label{identity}
e^z =  \frac{F(3z) - F(z) -\ln{3}}{F(2z) - F(z) - \ln{2}} -1
\end{equation}
allows us to recover the exponential function by standard algebraic simplification. The logarithm is then simply $\ln{z} = F(z) - e^z$. Once we have both $\exp$ and $\ln$, together with subtraction, we have \eqref{eml}, and from \cite{EML} we know that this is enough to evaluate all elementary functions. The constants used in \eqref{identity}, namely $1,2,3,\ln{2},\ln{3}$ are needed as written but are likely redundant. The complexity of \eqref{identity} in RPN form is K=23, far beyond direct enumeration reach, and it is not known whether it is the simplest possible construction of this kind. Nevertheless, compositions of rational operations and $F$ can in principle generate exact analytical formulas; Fig.~\ref{arch}, bottom panel, shows the basic layer. The matrix multiplication layer, including the output map, must be replaced by a rational function layer to allow the division required by \eqref{identity}, and complex numbers are required to use the EML-style reduction. The activation function is now \eqref{holly_molly}. The question whether such a network can indeed be optimized efficiently is beyond scope of this note.

\section{Conclusions}

The article discusses common misconceptions, previous knowledge, follow-up studies and emerging directions which came after the discovery of the EML operator \cite{EML}. What was missed by readers is that the EML itself, given by formula \eqref{eml} is probably only the first known member of a large family of operators and functions, which could open new ways of evaluating elementary functions, enhancing the abilities of machine learning. So far, no systematic theory has been created or exhaustive search done to reveal all its variants. One of the undiscovered variants could possibly enable exact analytical functions in an architecture nearly identical to existing neural network, keeping standard optimization procedures. In the optimistic variant, the only modification would be complex weights in place of real ones, and a new non-linear activation possibly resembling \eqref{holly_molly} in an algebraic way. If we are lucky,  the new activation will work in the full real domain, keeping the asymptotics of the ramp function (ReLU), i.e. zero for $x \to -\infty$ and $x$ for $x \to +\infty$. All EML-type nonlinearity would remain near zero and in imaginary direction. A natural candidate with the above properties is some combination of elementary hyperbolic functions. The use of complex numbers might not be necessary, as they can be traded for their matrix representation 
$
i \to \left(\begin{smallmatrix}0&-1\\1&0\end{smallmatrix}\right)
$
used in e.g. recent articles on quantum mechanics \cite{QM_no_i} without complex numbers \cite{QM_real}. However, if rational functions are indeed required to achieve the stated goal, it would be of no surprise to practitioners, as e.g. Pade approximation has been known to be a superior method for a long time.

\section*{Acknowledgments}

I would like to thank Henrik Klagges for the invitation to the TNG Big Techday conference in Munich, and the Faculty of Mathematics and Computer Science of the Jagiellonian University for support in the form of a Maple license. Computational resources were partially provided by Google Cloud Research Credits and the Polish National Science Centre MAESTRO Grant No.~2017/26/A/ST2/00530.

\bibliographystyle{plain}
\bibliography{EML-diversity}

\appendix

\section{Wolfram Mathematica implementation of a Goldstern-type single operator}

Let $\beta\colon\mathbb{Z}\to\mathbb{N}$ where $\mathbb{N} = \{0,1,2,\ldots\}$, be the standard pairing between integers and naturals
\begin{subequations}
\label{noconst}
\begin{equation}
\beta(m) =
\begin{cases}
2m & m \geq 0,\\
-2m-1 & m < 0,
\end{cases}
\qquad
\beta^{-1}(p) =
\begin{cases}
p/2 & p \;\text{even},\\
-(p+1)/2 & p \;\text{odd}.
\end{cases}
\end{equation}
Every positive integer factors uniquely as $2^l(2a+1)$. Applying this to
$\beta(\lfloor x \rfloor)+1$ assigns to every real $x$ a ``floor number''
$l(x) \in \mathbb{N}$ and a ``payload'' $u(x) \in \mathbb{R}$:
\begin{equation}
\beta(\lfloor x \rfloor)+1 = 2^{l(x)}(2a+1),
\qquad
u(x) = \beta^{-1}(a) + x - \lfloor x \rfloor .
\end{equation}
The map $x \mapsto \bigl(u(x),l(x)\bigr)$ is an explicit bijection
$\mathbb{R}\to\mathbb{R}\times\mathbb{N}$, with inverse
\begin{equation}
\rho(v,l) = \beta^{-1}\!\left[\, 2^l \bigl(2\beta(\lfloor v \rfloor)+1\bigr)-1 \,\right]
+ v - \lfloor v \rfloor,
\end{equation}
and $S(x) = \rho\bigl(u(x), l(x)+1\bigr)$ moves $x$ one floor up, keeping the payload.
The operator is
\begin{equation}
\label{diamondop}
x \diamond y =
\begin{cases}
S(x)            & \text{for} \;\; y = x,\\
\rho(x,0)       & \text{for} \;\; y = S(x),\\
u(x) + u(y)     & \text{for} \;\; l(x)=1,\; l(y)=0,\\
u(x) - u(y)     & \text{for} \;\; l(x)=2,\; l(y)=0,\\
u(x) \times u(y)& \text{for} \;\; l(x)=3,\; l(y)=0,\\
u(x) / u(y)     & \text{for} \;\; l(x)=4,\; l(y)=0,\; u(y)\neq 0,\\
x               & \text{otherwise.}
\end{cases}
\end{equation}
\end{subequations}
Writing $\sigma(t) = t \diamond t$ (one floor up) and
$\varepsilon(x) = x \diamond (x \diamond x)$ (encoding at floor zero) we obtain
\begin{eqnarray*}
x + y       &=& \sigma\bigl(\varepsilon(x)\bigr) \diamond \varepsilon(y), \\
x - y       &=& \sigma^{2}\bigl(\varepsilon(x)\bigr) \diamond \varepsilon(y), \\
x \times y  &=& \sigma^{3}\bigl(\varepsilon(x)\bigr) \diamond \varepsilon(y), \\
\frac{x}{y} &=& \sigma^{4}\bigl(\varepsilon(x)\bigr) \diamond \varepsilon(y),
\end{eqnarray*}
valid for all real $x,y$ (with $y \neq 0$ in the last line). No constant appears anywhere;
fully expanded, e.g.,
$$
x + y = \Bigl\{ \bigl[\, x \diamond (x \diamond x) \,\bigr] \diamond
                \bigl[\, x \diamond (x \diamond x) \,\bigr] \Bigr\}
        \diamond \bigl[\, y \diamond (y \diamond y) \,\bigr].
$$

Constants can also be generated from pure-$\diamond$ terms, e.g.
$$
0 = \Biggl\{ \Bigl( \bigl[\, x \diamond (x \diamond x) \,\bigr] \diamond
              \bigl[\, x \diamond (x \diamond x) \,\bigr] \Bigr) \diamond
     \Bigl( \bigl[\, x \diamond (x \diamond x) \,\bigr] \diamond
              \bigl[\, x \diamond (x \diamond x) \,\bigr] \Bigr) \Biggr\}
    \diamond \bigl[\, x \diamond (x \diamond x) \,\bigr].
$$

Below you can find a working implementation of the above procedure in Wolfram Mathematica. 

\begin{lstlisting}[language=Mathematica, basicstyle=\ttfamily\footnotesize, breaklines=true]
(* Pairing integers to naturals *)
beta[m_Integer]  := If[m >= 0, 2 m, -2 m - 1];
ibeta[p_Integer] := If[EvenQ[p], p/2, -(p + 1)/2];

(* Floor number and payload: x <-> (val, level) is a bijection R <-> R x N *)
level[x_] := IntegerExponent[beta[Floor[x]] + 1, 2];
val[x_]   := Module[{p = beta[Floor[x]] + 1, l},
               l = IntegerExponent[p, 2];
               ibeta[(p/2^l - 1)/2] + (x - Floor[x])];
rho[v_, l_Integer] := ibeta[2^l (2 beta[Floor[v]] + 1) - 1] + (v - Floor[v]);
S[x_] := rho[val[x], level[x] + 1];    (* same payload, one floor up *)

(* THE single operator *)
op[x_, y_] := Which[
   y == x,    S[x],                    (* climb *)
   y == S[x], rho[x, 0],               (* encode *)
   1 <= level[x] <= 4 && level[y] == 0,
     Switch[level[x],
       1, val[x] + val[y],
       2, val[x] - val[y],
       3, val[x] val[y],
       4, If[val[y] == 0, x, val[x]/val[y]]],
   True, x];

(* Pure-op terms for the four operations *)
s[x_]   := op[x, x];
enc[x_] := op[x, s[x]];
plus[x_, y_]   := op[s[enc[x]], enc[y]];
minus[x_, y_]  := op[s[s[enc[x]]], enc[y]];
times[x_, y_]  := op[s[s[s[enc[x]]]], enc[y]];
divide[x_, y_] := op[s[s[s[s[enc[x]]]]], enc[y]];

(* Examples *)

plus=op[op[op[x, op[x, x]], op[x, op[x, x]]], op[y, op[y, y]]] (* x+y *)
zero=op[op[op[op[x, op[x, x]], op[x, op[x, x]]], op[op[x, op[x, x]], op[x, op[x, x]]]], 
op[x, op[x, x]]] (* zero *)
(* Some values are required to resolve conditionals *)
plus /. {x->EulerGamma, y->Glaisher}
zero /. x->Khinchin
\end{lstlisting}

\end{document}